\documentclass[aps,prb,groupedaddress,twocolumn,superscriptaddress]{revtex4}
\usepackage{graphicx}
\usepackage{epsfig}
\usepackage{ulem}
\usepackage{color}
\usepackage{bm}
\usepackage{mathrsfs}
\usepackage{dcolumn}
\usepackage{amsmath, amssymb}

\begin{document}

\title{Theory for electric dipole superconductivity with an application for bilayer excitons}

\author{Qing-Dong Jiang}
\affiliation{International Center for Quantum Materials, School of Physics, Peking University, Beijing 100871, P.R. China}
\author{Zhi-qiang Bao}
\affiliation{Institute of Physics, Chinese Academy of Sciences, Beijing 100190, China}
\author{Qing-Feng Sun}
\email[]{sunqf@pku.edu.cn}
\affiliation{International Center for Quantum Materials, School of Physics, Peking University, Beijing 100871, P.R. China}
\affiliation{Collaborative Innovation Center of Quantum Matter, Beijing 100871, P.R. China}
\author{X. C. Xie}
\email[]{xcxie@pku.edu.cn}
\affiliation{International Center for Quantum Materials, School of Physics, Peking University, Beijing 100871, P.R. China}
\affiliation{Collaborative Innovation Center of Quantum Matter, Beijing 100871, P.R. China}

\date{\today}

\maketitle

\textbf{Exciton superfluid is a macroscopic quantum phenomenon in
which large quantities of excitons undergo the Bose-Einstein
condensation. Recently, exciton superfluid has been widely studied
in various bilayer systems. However, experimental measurements only
provide indirect evidence for the existence of exciton superfluid.
In this article, by viewing the exciton in a bilayer system as an
electric dipole, we provide a general theory for the electric dipole
superconductivity, and derive the London-type and
Ginzburg-Landau-type equations for the electric dipole
superconductors. By using these equations, we discover the
Meissner-type effect and the electric dipole current Josephson
effect. These effects can provide direct evidence for the formation
of the exciton superfluid state in bilayer systems and pave new ways
to drive an electric dipole current.}\\

Since the idea of excitonic condensation was proposed about fifty
years ago\cite{Keldysh1965,Lozovik1975}, exciton systems have
attracted a lot of interest. With the development of micromachining
technology in the last two decades, high-quality bilayer exciton
systems can be fabricated in the laboratories, in which one layer
hosts electrons and the other layer hosts
holes.\cite{science1368,eisenstein2004} Many new interaction
phenomena have been experimentally reported in the bilayer exciton
systems,\cite{kellogg2004vanishing,Naturephy799,spielman2000resonantly,
tutuc2004counterflow,nandi2012exciton,gorbachev2012strong,Science99,titov2013giant,
Nature409,Nature584} including the vanishing Hall resistance for
each layer\cite{kellogg2004vanishing}, the resonantly enhanced
zero-bias inter-layer tunneling
phenomenon\cite{spielman2000resonantly}, the large bilayer
counterflow conductivity\cite{tutuc2004counterflow}, the Coulomb
drag\cite{nandi2012exciton,gorbachev2012strong,Science99,titov2013giant},
etc. These phenomena strongly imply the formation of the exciton
condensate superfluid state, in which many excitons crowd into the
ground state. However, because the exciton is charge neutral, there
are still no effective methods to directly confirm the formation of
the superfluid state. Thus, whether the superfluid state really
forms is still unclear.

Before any further discussion, we need first to point out the
specificity of excitons in bilayer systems. Because the electrons
and holes are separated in space and bound with each other by the
Coulomb interaction, the exciton in a bilayer system can be seen as
a charge neutral electric dipole (as shown in Fig. \ref{fig:1}a). On
the other hand, superconductivity has been one of the central
subjects in physics. The superconductor state has several
fascinating properties, such as zero resistance\cite{onnes}, the
Meissner effect\cite{meissner}, the Josephson
effect\cite{Josephson1962}, and so on, which have many applications
nowadays\cite{gennes1989superconductivity}. It is now well known
that the superconductor is the condensate superfluid state of the
Cooper pairs\cite{bardeen1957theory}, which can be viewed as
electric monopoles. In other words, the superconductor state is the
electric monopole condensated superfluid state. Thus, it is natural
to ask whether the electric dipole superfluid state possesses many
similar fascinating properties, just like its counterpart, the
electric monopole superfluid state.

In this article, we will construct a general theory of electric
dipole superconductivity under an external electromagnetic field,
and apply this theory to the bilayer exciton systems, revealing the
basic characteristics of the electric dipole superconductors. Apart
from the bilayer exciton systems, the electric dipole superconductor
may also exist in other two or three dimensional systems, e.g., the
Bose-Einstein condensate of ultracold polar
molecules\cite{phystoday64-27,nature464-1324,science322-231}. In
fact, the ultracold polar molecules have been successfully produced
in the laboratory over the past
decade\cite{phystoday64-27,nature464-1324,science322-231}. In
addition, a new quantum state was proposed
recently\cite{Sun2011,Sun2013,Bao2013}, a magnetic dipole
superconductor named the spin superconductor. Both electric and
magnetic dipole superconductors contain some similar properties.
Below we first derive the London-type and Ginzburg-Landau-type
(GL-type) equations of the electric dipole superconductor. These
equations can be applied to all electric dipole superconductors
independent of specific systems and we apply them to study various
physical properties of the electric dipole superconductor. By using
these equations, we find that the Meissner-type effect and the
Josephson effect of the electric dipole current. With the
Meissner-type effect, a non-uniform external magnetic field can
cause a super electric dipole current in an electric dipole
superconductor, and a super electric dipole current can generate a
magnetic field that is against the spatial variation of the external
magnetic field. Considering the bilayer exciton systems, we show
that the magnetic field induced by the super dipole current is
measurable by today's technology. We also show that the frequency in
the AC Josephson effect of the electric dipole current is equal to
that of the AC Josephson effect in the normal superconductor. These
new effects discovered in this work can not only provide direct
evidence for the existence of the exciton condensate superfluid
state in the bilayer systems, but also pave new ways to drive an
electric dipole current.

\begin{figure}[!htb]
\includegraphics[height=5.4cm, width=8cm, angle=0]{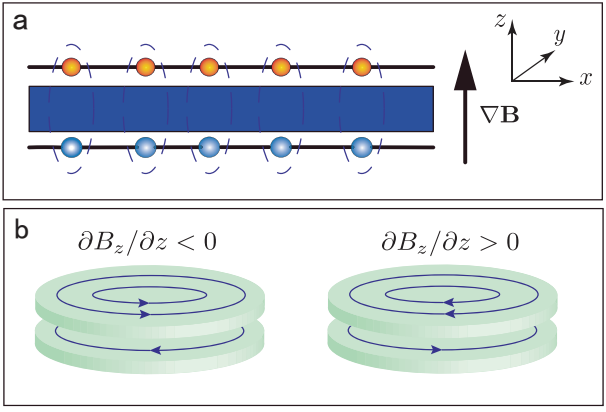}
\caption{ \textbf{A side view of the exciton in bilayer system and
the induced supercurrent by a non-uniform magnetic field.} (\textbf
{a}) The top and bottom layers host holes and electrons
respectively, and the middle blue block stands for the interlayer
barrier which prevents tunneling between the two layers. (\textbf
{b}) The left (right) panel shows the induced super electric dipole
current for $\partial B_z/\partial z<0$ ($\partial B_z/\partial
z>0$). The arrows on the blue lines denote the direction of positive
charge flow in each layer.\label{fig:1}}
\end{figure}

\begin{center}
\textbf{Results}
\end{center}

\begin{center}
\textbf{London-type equations of the electric dipole superconductor}
\end{center}

Considering a bosonic electric dipole condensate superfluid state,
namely the electric dipole superconductor, under an external
electric field, the force on the electric dipole $\bold{p}_0$ is
\begin{eqnarray}\label{eq:1}
\bold{F}=(\bold{p}_0 \cdot\nabla)\bold{E}=m^*\frac{d\bold v}{dt}
\end{eqnarray}
which accelerates the dipole. Here, $\bold E$ is the electric field,
$m^*$ is the effective mass of the dipole and $\bold v$ is its
velocity. The moving electric dipole induces an electric dipole
current, which can be described by a tensor $\mathbb{J}_{\bold
p}=n\bold{p_0}\bold{v}$, where $n$ is the dipole density. It should
be pointed out that, in this work the direction of the electric
dipole vector is fixed in the $z$-direction (as is the case in a
bilayer exciton system), thus we only need one vector $\bold
{J_p}=np_0\bold{v}$ to describe the super dipole current density.
The derivative of $\bold{J_p}$ with respect to time is
\begin{eqnarray}\label{eq:2}
\frac{d{\bold {J_p}}}{dt}=\beta(\hat p_0 \cdot \nabla )\bold E,
\end{eqnarray}
where $\beta=n p_0^2/m^*$ and $\hat p_0 ={\bold e}_z$ is the unit
vector in the direction of $\bold p_0$. We can see that, just as
$\bold E$ accelerates the super electric
current\cite{gennes1989superconductivity}, $(\hat
p_0\cdot\nabla)\bold E$ accelerates the super electric dipole
current. Taking curl on both sides of the equation (\ref{eq:2}) and
using the Maxwell equation\cite{griffiths} $\nabla\times\bold
E=-\partial\bold B/ \partial t$, we have $\frac{\partial}{\partial
t}(\nabla\times\bold J_p)=-\frac{\partial}{\partial t}[\beta(\hat
p_0\cdot\nabla)\bold B]$. Integrating over the time $t$, we get the
London-type equation for $\bold J_p$,
\begin{eqnarray}\label{eq:3}
\nabla \times{\bold J_p}=-\beta (\hat p_0\cdot \bold\nabla)\bold B,
\end{eqnarray}
where the integral constant is taken to be zero due to the
requirement of thermodynamic equilibrium. Equations (\ref{eq:2}) and
(\ref{eq:3}) play similar roles as the London equations for normal
superconductors\cite{gennes1989superconductivity}, so we call them
the first and the second London-type equations for the electric
dipole superconductor.

Equation (\ref{eq:3}) implies that the gradient of a magnetic field
${\bold B}$ will induce a super electric dipole current. As is shown
in Fig. \ref{fig:1}b, if the gradient of magnetic field $\partial
B_z/\partial z<0$, the super dipole current flows in the
counterclockwise direction (left panel of Fig.\ref{fig:1}b); if
$\partial B_z/\partial z>0$, the super dipole current flows in the
clockwise direction (right panel of Fig.\ref{fig:1}b). In addition,
the super dipole current can also have a feedback for an external
magnetic field. The magnetic field generated by a moving electric
dipole $\bold p_0$ with velocity $\bold v$ is equivalent to that
generated by a static magnetic moment $\bold m=-\bold v\times\bold
p_0$.\cite{griffiths,Magneticdipole} As a consequence, the magnetic
field induced by the super dipole current $\bold J_p$ is equivalent
to that induced by the static magnetic moment distribution
(magnetization) $\bold M=n \bold m=-\bold J_p\times \hat p_0$. In
materials, the last Maxwell equation takes the form
$\nabla\times\bold B=\mu_0(\bold j_f+\nabla\times\bold M+{\partial
\bold D}/{\partial t})$,\cite{griffiths} where $\bold j_f$ stands
for the free electric current and $\bold D$ is the effective
electric field. In the equilibrium case, ${\partial \bold
D}/{\partial t} =0$ and with no free electric current present, only
the super dipole current exists, so we obtain the magnetic field
equation in the electric dipole superconductor:
\begin{eqnarray}\label{eq:4}
\nabla\times\bold B=-\mu_0 \nabla\times(\bold J_p\times \hat p_0).
\end{eqnarray}

The London-type equation (\ref{eq:3}) and the magnetic field
equation (\ref{eq:4}) govern the magnetic field and the super dipole
current in an electric dipole superconductor. An alternative set of
equations that are equivalent to equations (\ref{eq:3}) and
(\ref{eq:4}) are given in Supplementary Note 1. From equations
(\ref{eq:3}) and (\ref{eq:4}), we can obtain the Meissner-type
effect against the spatial variation of an external magnetic field,
which will be studied below. We can see the effect of equation
(\ref{eq:3}) by considering a massless Dirac particle for which the
factor $\beta\rightarrow\infty$. In this case, the spatial variation
of the total magnetic field $\partial_z B_z$ has to vanish
everywhere inside an electric dipole superconductor in order to
satisfy the equation (\ref{eq:3}). This means that $\partial_z B_z$
is completely screened out.

\begin{center}
\textbf{Ginzburg-Landau-type equations of the electric dipole
superconductor}
\end{center}

Since the electric dipole condensate is a
macroscopical quantum state, we can use a quasi-wave function (or
the order parameter) $\psi(\bold r)$ to describe it. Then its free
energy can be written as $F_s$=$\int_V f_s d\bold r$, where $f_s$ is
the free energy density. In analogy with the superconductor, $f_s$
can be expressed as:
\begin{multline}\label{eq:5}
f_s =f_n +\alpha(T)|\psi(\bold r)|^2+\frac{\beta(T)}{2}|\psi(\bold r)|^4+ \\
\frac{|(\hat{\bold p} + \bold p_0\times \bold B)\psi(\bold r)|^2}{2m^*}+\frac{|\bold B|^2}{2\mu_0} ,
\end{multline}
where $f_n$ is the density of free energy in normal state and the
momentum operator ${\bold {\hat p}}=-i\hbar\nabla$. The two terms
$\alpha(T)|\psi(\bold r)|^2$ and $\beta(T)|\psi(\bold r)|^4/2$ are
the lower order terms in the series expansion of the free energy
$f_s$, which have similar meanings as those in the normal
superconductor.\cite{gennes1989superconductivity,GLequation}
Particularly, the gauge invariant term $|(\bold p + \bold
p_0\times\bold B)\psi(\bold r)|^2/2m^*$ can be viewed as the kinetic
energy of the electric dipole superconductor (see Supplementary
Note 2). Substitute ${\mathbf B}=\nabla\times{\mathbf A}$ and
minimize the free energy with respect to $\psi^*$ and the magnetic
vector potential $\bold A$ respectively, we get (see Supplementary
Note 3)
\begin{eqnarray}
&&\mbox{}\hspace{-10mm}\alpha(T)\psi+\beta(T)|\psi|^2\psi+
\frac{[\hat{\bold p}+\bold p_0\times \bold B]^2\psi}{2m^*}=0, \label{eq:6}\\
&&\mbox{}\hspace{-10mm}\nabla\times\bold B=-\mu_0 \nabla\times(\bold J_p\times \hat p_0),\label{eq:7}
\end{eqnarray}
where
\begin{eqnarray}\label{eq:8}
\bold J_p=\frac{p_0}{2
m^*}[i\hbar(\psi\nabla\psi^*-\psi^*\nabla\psi)+2\bold p_0\times\bold
B\,|\psi|^2] .
\end{eqnarray}
Equations (\ref{eq:6}) and (\ref{eq:8}) are the first and the second
GL-type equations, respectively. They provide a full
phenomenological description of the dipole superconductor. Since the
velocity operator is $\bold {\hat v}=(\bold {\hat p}+\bold
p_0\times\bold B)/m^*$, the super electric dipole current can be
expressed as $p_0 Re(\psi^*\bold{\hat v}\psi)=\frac{p_0}{2
m^*}[i\hbar(\psi\nabla\psi^*-\psi^*\nabla\psi)+2\bold p_0\times\bold
B\,|\psi|^2]$. Comparing the expression $p_0 Re(\psi^*\bold{\hat
v}\psi)$ and equation (\ref{eq:8}), we find that $\bold J_p$ is
exactly the super dipole current density. It should be noted that
equation (\ref{eq:7}) is the same as the magnetic field equation
(\ref{eq:4}). It indicates that the GL-type theory gives a more
general result. Next we derive the London-type equations from the
GL-type equations. The order parameter $\psi(\bold r)$ can be
written as $|\psi(\bold r)|e^{i\theta(\bold r)}$, where $|\psi(\bold
r)|^2$ is proportional to the density of dipoles $n$ and $\theta$
represents the phase. For simplicity, we assume the amplitude
$|\psi|$ is the same everywhere in the dipole superconductor,
whereas the phase $\theta(\bold r)$ are allowed to change in order
to account for the super dipole current. Substitute
$\psi=|\psi(\bold r)|e^{i\theta(\bold r)}$ into equation
(\ref{eq:8}), we get $\bold J_p=\frac{p_0 n}{m^*}(\hbar
\nabla\theta+\bold p_0\times\bold B)$. Furthermore, if we take curl
on both sides, the London-type equation (\ref{eq:3}) is recovered.
It indicates that the London-type equations can be obtained from the
GL-type equations, which shows the validity and consistency of our
theory.

\begin{figure*}[!htb]
\includegraphics[height=3.9cm, width=16.9cm, angle=0]{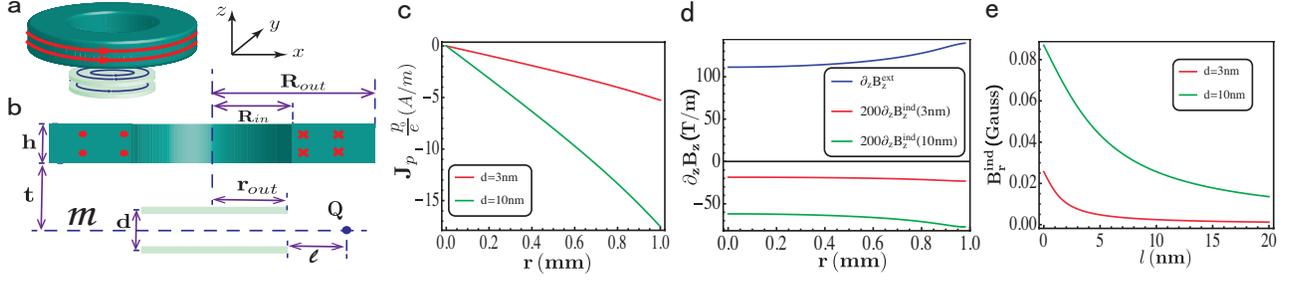}
\caption{ \textbf{Meissner-type effect of the dipole superconductor.}
(\textbf{a,b}) The schematic diagram of the device consisting of a
cylindrical hollow conductor and a bilayer exciton system (the
dipole superconductor), and the cross section of it. (\textbf{c,d})
The induced super dipole current ${\mathbf J}_{\bold p}$ and the
spatial variation of the induced magnetic field $\partial_z
B_z^{ind}$ in the middle plane m versus the coordinate $r$.
(\textbf{e}) The induced magnetic field $B_r^{ind}$ versus the
distance $l$ for thickness $d=3\,nm$ and $d=10\,nm$.\label{fig:2}}
\end{figure*}

\begin{center}
\textbf{Meissner-type effect of the electric dipole superconductor}
\end{center}

In the following, we will use the London-type equations to analyse the
Meissner-type effect. To begin with, we consider a two-dimensional
circular dipole superconductor with the radius $r_{out}$ located in
a non-uniform external magnetic field $\bold B$ created by a
cylindrical hollow conductor with the inner (outer) radius $R_{in}$
($R_{out}$) and the height $h$ (shown in Fig. \ref{fig:2}a). The
distance between the cylindrical hollow conductor and the dipole
superconductor is $t$. Fig. \ref{fig:2}b depicts the cross-section
of the device. A uniform electric current along the azimuthal
direction in the hollow conductor creates a non-uniform magnetic
field with a gradient $\partial_{z}B_z^{ext}$ (see Supplementary
Note 4), which can induce a super dipole current in the
electric dipole superconductor. Substitute $\partial_z B_z^{ext}$
into the London-type equation (\ref{eq:3}), considering the
rotational symmetry of the whole device and $\nabla \cdot {\bold
J}_p =0$, we can obtain the super electric dipole current density
$J_p$ (flowing in the azimuthal direction):
\begin{eqnarray}\label{eq:9}
J_p(r)=\frac{-\beta}{r}\int_0^r r^{\prime}dr^{\prime}[\partial_z B_z^{ext}(r^{\prime},z)].
\end{eqnarray}

In the following, we consider that the electric dipole
superconductor is the bilayer exciton system (see Fig.
\ref{fig:1}a). Then the super dipole current can be viewed as a
counterflow electric current in the bilayer as shown in
Fig.\ref{fig:1}b and thereby induces a magnetic field $\bold
B^{ind}$. Since the counterflow current in bilayer and the
rotational symmetry about the $z$ axis, the induced magnetic field
$\bold B^{ind}$ in middle plane $m$ only has the nonzero r-component
$B^{ind}_r$. Although the z-component $B^{ind}_z =0$, the gradient
$\partial_z B_z^{ind}$ does not vanish. Fig. 2d and 2e show
$\partial_z B_z^{ind}$ and $B^{ind}_r$ in plane $m$ which can easily
be calculated from the Biot-Savart law (see Supplementary Note 5).

In the calculation, we take cylindrical hollow conductor sizes as
$R_{in}=1mm$, $R_{out}=1\,cm$ and $h=1.5\,cm$. The current density
in conductor $j=10^8 A/m^2$, which generates the non-uniform
external magnetic field ${\bold B}^{ext}$. The bilayer exciton
specimen, the electric dipole superconductor, is below the conductor
with $t=0.1\, mm$ and the radius $r_{out}=1mm$. Here the specimen is
just in the hollow region of the conductor, because $\partial_z
B_z^{ext}$ is relatively large there (see Supplementary
Note 4). The two-dimensional carrier density in each layer $n$
is chosen $10^{12}cm^{-2}$ and the effective mass of exciton
$m^*=0.01 m_e$ with the electron mass $m_e$. Fig. \ref{fig:2}c,
\ref{fig:2}d and \ref{fig:2}e show respectively the induced super
dipole current density $J_p$, the induced magnetic field variation
$\partial_z B_z^{ind}$ and $B_r^{ind}$ versus radius $r$ for bilayer
thickness $d=3nm$ and $d=10 nm$. A quite large $J_p$ is induced near
the edge of the specimen, in which the corresponding electric
current density in each layer near the edge is about $15 A/m$. From
Fig. \ref{fig:2}d, we find that $\partial_z B_z^{ind}(r)$
counteracts the variation $\partial_z B_z^{ext}(r)$. This is a
Meissner-type effect in the dipole superconductor against a spatial
variation of a magnetic field. Notice that it is not against the
magnetic field. This is the main difference between the dipole
superconductor and (monopole) superconductor. Also notice in Fig.
\ref{fig:2}d, $\partial_z B_z^{ind}(r)$ is much smaller than
$\partial_z B_z^{ext}(r)$ because the thickness $d$ of the dipole
superconductor is very small now. If for the thick dipole
superconductor or for very small $m^*$, $\partial_z B_z^{ind}(r)$
can almost be of the  same value as $\partial_z B_z^{ext}(r)$, then
the spatial variation of the total magnetic field vanishes inside of
the dipole superconductor. Fig. \ref{fig:2}e shows the induced
magnetic field $B_r^{ind}$ by the super dipole current, which can
reach about 0.05 Gauss. This magnetic field can be accurately
detected by the today's technology. In addition, a recent work has
successfully used the SQUID to detect a tiny edge current (around
$0.5 A/m$) in the Hall specimen\cite{nowack2013imaging}. In our
case, the edge current density is around $10 A/m$, so it should be
detectable using the same method.

\begin{figure*}[!htb]
\includegraphics[height=3.9cm, width=16.9cm, angle=0]{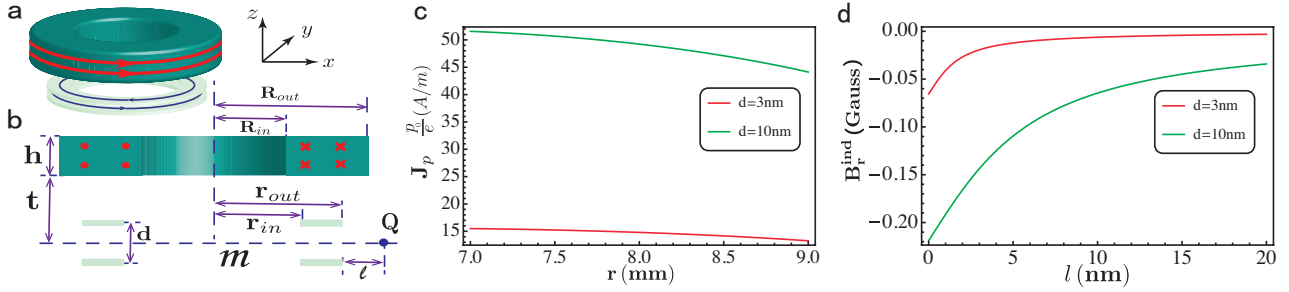}
\caption{ \textbf{The proposed device for detection of the zero
dipole resistance.} (\textbf{a,b}) The schematic diagram of the
proposed device and the cross section of it. (\textbf{c,d}) The
excited super electric dipole current and its magnetic field $
B_r^{ind}$ with the sizes of the dipole superconductor $r_{in}=7mm$
and $r_{out}=9mm$. The other parameters are the same as in Fig.
\ref{fig:2}.\label{fig:3}}
\end{figure*}

\begin{center}
\textbf{The detection of the zero dipole resistance}
\end{center}

The most remarkable phenomenon of the Bose-Einstein condensate macroscopic
quantum system is superfluid, e.g. the zero resistance phenomenon of
the (monopole)
superconductor\cite{onnes,gennes1989superconductivity}. For the
dipole superconductor, the dipole resistance is zero, i.e. the
electric dipole can flow without dissipation. In the following, we
suggest a method to detect the zero dipole resistance.

From the first London-type equation (\ref{eq:2}), we know that a
variation of an electric field $\partial_z \bold E$ can excite a
super dipole current. This excited super dipole current will
maintain for a very long time if the dipole resistance is zero. Now,
consider an annular dipole superconductor specimen. This annular
specimen is placed below the cylindrical hollow conductor (see Fig.
\ref{fig:3}a and Fig. \ref{fig:3}b), and there $\partial_z
B_z^{ext}$ is relatively small and $\partial_z A_{\theta}^{ext}$ is
quite large\cite{simpson} (see Supplementary Note 4).
First, let the hollow conductor have an azimuthal electric current
$j$ and then cool the specimen into the dipole superconductor state.
Next, we abruptly turn off the current $j$ in the conductor. In this
process, an azimuthal dipole current ${J_p}$ will be excited. From
the equation (\ref{eq:2}) and $\bold E=-\frac{\partial \bold
A}{\partial t}$, we have $d J_p/dt = \beta
\partial_z E_{\theta}^{ext} = -\frac{\partial}{\partial t}[\beta \partial_z
A_{\theta}^{ext}]$. Integrating over the time $t$, we obtain the excited
super dipole current:
\begin{eqnarray}\label{eq:10}
J_p =  \beta \partial_z A_{\theta}^{ext},
\end{eqnarray}
where the vector potential $A_{\theta}^{ext}$ is that before the
current in the conductor was turned off. Here we have used
$A_{\theta}^{ext}=0$ after the current turned off and $J_p$ is
almost zero at the beginning. Fig. \ref{fig:3}c and \ref{fig:3}d
show the excited super dipole current $J_p$ and the magnetic field
$B_r^{ind}$ induced by $J_p$. From which we find that $J_p$ and
$B_r^{ind}$ are quite large. Moreover, this excited super dipole
current $J_p$ does not decay for a very long time because of the
zero dipole resistance. So one can measure the non-decayness of
$J_p$ to confirm the zero dipole resistance.

\begin{figure*}[!htb]
\includegraphics[height=3cm, width=12.9cm, angle=0]{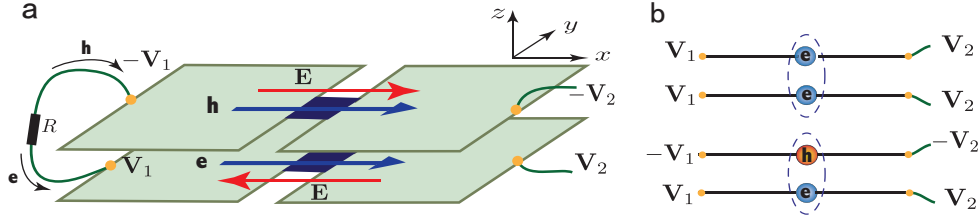}
\caption{\textbf{The electric dipole current Josephson junction.}
(\textbf{a}) The schematic diagram for the device of dipole current
Josephson junction, the bilayer exciton system-insulator-bilayer
exciton system junction. The left sides of the two layers are
connected by a wire, and the right sides of the two layers are
connected with voltages $V_2$ and $-V_2$, respectively. The red
arrows denote the direction of the electric fields in the top layer
and bottom layer, respectively, while the blue arrows represent the
flowing direction of holes (electrons) in the top (bottom) layer.
(\textbf{b}) The comparison between (monopole) superconductivity and
dipole superconductivity.\label{fig:4}}
\end{figure*}

\begin{center}
\textbf{DC and AC electric dipole current Josephson effect}
\end{center}

The Josephson effect is another highlight of the
superconductor\cite{Josephson1962}. Below we use the GL-type
equations of the electric dipole superconductor to discuss the
dipole current Josephson effect in a dipole
superconductor-insulator-dipole superconductor junction. From the
GL-type equations (\ref{eq:6}) and (\ref{eq:8}) we can see that if
${\mathbf B}=0$, they are the same as the GL equations of the
superconductor when ${\mathbf A}=0$. Therefore, the DC Josephson
effect of the dipole superconductor and the (monopole)
superconductor are similar, {\it i.e.}, the super electric dipole
current is $j=j_{0}\sin\gamma_{0}$, where
$\gamma_{0}=\gamma_{1}-\gamma_{2}$. $j_{0}$ is the Josephson
critical super dipole current, and $\gamma_{1}$, $\gamma_{2}$ are
phases of the dipole superconductors.

Now we study the AC dipole current Josephson effect and consider an
electric field variation $\partial_z E_x$. From the first GL-type
equation (\ref{eq:6}), we can get the change of the phase when the
super dipole current (in $x$-direction) passes through the Josephson
junction. Its expression is
$\gamma=\gamma_{0}+\frac{1}{\hbar}\int_{1}^{2}({\mathbf
p}_{0}\times{\mathbf B})\cdot{\mathbf e}_{x}dx$. Taking the
derivative with respect to time, we get
$\frac{\partial\gamma}{\partial
t}=\frac{p_{0}}{\hbar}\int_{1}^{2}({\mathbf
e}_{z}\times\frac{\partial{\mathbf B}}{\partial t})\cdot{\mathbf
e}_{x}dx$. Substituting $\frac{\partial{\mathbf B}}{\partial
t}=-\nabla\times{\mathbf E}$, we have $\gamma=\gamma_0 -\omega_0 t$
with $\omega_0 =-\frac{p_0 }{\hbar}\int_{1}^{2}\partial_z E_xdx$. As
a result, the super dipole current can be written as $j=j_0 \sin
(\gamma-\omega_0 t)$. It shows that the dipole current is an
alternating current, although the electric field spatial variation
is time-independent. We can compare this with the (monopole)
superconductor. For the superconductor, a time-independent electric
field (or bias) can drive an AC Josephson current. Now a spatial
variation of an electric field drives an AC dipole Josephson
current.

Next, we consider that the dipole superconductor is the bilayer
exciton system. Fig. \ref{fig:4}a shows the schematic diagram for
the device of dipole current Josephson junction. A thin wire
connects the left sides of the two layers, which enables the current
to flow between them. Then, if we apply the voltages $-V_2$ and
$V_2$ to the right sides between the bilayer, it establishes an
electric field in $-x$ ($x$) direction in the bottom (top) layer
(see Fig. \ref{fig:4}a). This means that a spatial variation of
electric field, $\partial_z E_x$, is added on junction, so an AC
dipole Josephson current is driven and an alternating electric
current emerges in the external circuit, although only a DC bias is
added. For the bilayer exciton system, $p_0=ed$ and $\partial_z E_x
= 2E_x/d$, so the frequency $\omega_0 =-\frac{2 e d
}{d\hbar}\int_{1}^{2} E_xdx = \frac{2e}{\hbar} (V_2-V_1)$. This
frequency is the same with that of the AC Josephson effect of the
superconductor\cite{Josephson1962}. The reason is as follows. In the
electric current Josephson effect, two electrons form a Cooper pair,
and the Cooper pair moves in response to external voltages. In the
dipole current Josephson effect in the bilayer system, however, a
pair consists of an electron and a hole, which moves in response to
the counter voltages. The difference between them is shown in Fig.
\ref{fig:4}b. It can be seen that the pairs feel the same electric
forces in the superconductor and the dipole superconductor. Thus the
frequencies of the alternating current should be the same in the two
cases. The fact that the results are indeed exactly the same shows
that these results of the dipole superconductor are reasonable and
credible. In addition, since its frequency is the same as that of
the superconductor, it is not difficult to measure in an experiment
because the latter has been observed for a long time. As a result,
detecting the electric dipole current Josephson effect is another
feasible method to verify the formation of the dipole
superconductor.

\begin{center}
\textbf{Discussion}
\end{center}

In conclusion, we view the exciton condensate superfluid state
in bilayer electron system as an electric dipole superconductor
state. Then, from the properties of the electric dipoles in an
external electromagnetic field, we derive the London-type and the
GL-type equations for an electric dipole superconductor. These
equations are universal to all dipole superconductors and can also
be used to study various properties of a dipole superconductor. By
using these equations, we discover the Meissner-type effect against
the spatial variation of the magnetic field and the DC and AC dipole
current Josephson effects, and also suggest a method to detect the
zero dipole resistance. These new effects discovered in this work
can not only provide direct evidence for the existence of the
exciton superfluid in bilayer electron systems, but also pave new
ways to drive an electric dipole current.

\begin{center}
\textbf{Acknowledgments}
\end{center}
This work was financially supported by NBRP of China (2012CB921303
and 2012CB821402) and NSF-China under Grants Nos. 11274364 and
91221302.

\begin{center}
\textbf{Author contributions}
\end{center}
Q.D.J., Z.Q.B., Q.F.S., and X.C.X. performed the calculations,
discussed the results and wrote this manuscript together. Q.D.J.
and Z.Q.B. contributed equally to this work.

\begin{center}
\textbf{Additional information}
\end{center}

\textbf{Supplementary Information} accompanies this paper at
http://www.nature.com/naturecommunications

\textbf{Correspondence and requests} for materials
should be addressed to Q.F.S. or X.C.X.

\textbf{Competing financial interests:} The authors declare
no competing financial interests.

\clearpage
\begin{widetext}
\appendix
\section{Supplementary Information for ``Theory for electric dipole superconductivity with an application for bilayer excitons"}

\begin{center}
Qing-Dong Jiang, Zhi-qiang Bao, Qing-feng Sun, and X. C. Xie
\end{center}

\date{\today}
\maketitle

\begin{center}
\textbf{Supplementary Note 1: An alternative set of equations for the electric dipole superconductor.}
\end{center}

Starting from Eqs. (3) and (4) in the main text, we can get an
alternative set of equations to describe the electric dipole
superconductor in the steady state. First of all, make some
simplification on Eq. (4) in the main text, i.e,
\begin{eqnarray}\label{eq:s1}
\begin{split}
\nabla\times\bold B=&-\mu_0\nabla\times(\bold J_p\times\hat p_0)\\=&-\mu_0(\hat p_0\cdot\nabla)\bold J_p-\mu_0\hat p_0(\nabla\cdot\bold J_p)\\=&-\mu_0(\hat p_0\cdot\nabla)\bold J_p.
\end{split}
\end{eqnarray}
The term $\nabla\cdot\bold J_p$ is taken to be zero in the steady
state because of the conservation of the super electric dipole
current in material. By taking curl on both sides of the London-type
equation (3) in the main text, one has $\nabla\times(\nabla\times\bold
J_p)=-\beta(\hat p_0\cdot\nabla)(\nabla\times\bold B)$. Combining
this equation and Eq. (\ref{eq:s1}), we obtain the equation
for the super electric dipole current density
\begin{eqnarray}\label{eq:s2}
\nabla\times(\nabla\times\bold J_p)=\mu_0 \beta (\hat p_0\cdot\nabla)^2\bold J_p\label{eq:s2},
\end{eqnarray}
where the characteristic dimensionless ratio $\mu_0 \beta=\mu_0 n
p_0^2/m^*$ governs the screening strength. If we take curl on both
sides of Eq. (\ref{eq:s1}), it transforms into
$\nabla\times(\nabla\times\bold B)=-\mu_0(\hat
p_0\cdot\nabla)(\nabla\times\bold J_p)$. Combine this equation and
the London-type equation (3) in the main text, then we get the
equation for magnetic field
\begin{eqnarray}\label{eq:s3}
\nabla\times(\nabla\times\bold B)=\mu_0 \beta (\hat p_0\cdot\nabla)^2\bold B.
\end{eqnarray}
Now, we have obtained an alternative set of Eqs. (\ref{eq:s2})
and (\ref{eq:s3}) describing the magnetic field and the super
electric dipole current separately in the electric dipole
superconductor.

\vspace{5mm}

\begin{center}
\textbf{Supplementary Note 2: The Hamiltonian of a moving electric dipole and the
kinetic energy term of the electric dipole superconductor}
\end{center}

An electric dipole moving with velocity $\bold v$ in magnetic field
$\bold B$ can feel an electric field $\bold E^{\prime}=\bold
v\times\bold B$, and the corresponding energy is $-\bold
p_0\cdot\bold E^{\prime}=-\bold p_0\cdot(\bold v\times\bold B)=\bold
v\cdot(\bold p_0\times \bold B)$.\cite{griffiths,Magneticdipole}
Therefore, the Lagrangian of this electric dipole is
$\mathscr{L}=\frac{1}{2}m^*\bold v^2-\bold v\cdot(\bold p_0\times
\bold B)$, where $m^*$ is the effect mass of the electric dipole.
The canonical momentum is $\bold p=\frac{\partial \mathscr
L}{\partial \bold v}=m^*\bold v-\bold p_0\times \bold B$. Thus the
Hamiltonian of a moving electric dipole in a magnetic field is
\begin{eqnarray}\label{eq:s4}
\mathscr H =\bold {\hat p}\cdot\bold v-\mathscr L=\frac{[\hat{\bold p} + \bold p_0\times \bold B]^2}{2m^*}.
\end{eqnarray}
The term $\bold p_0\times \bold B$ is analogous to the term $e\bold
A/c$ for an electron in a magnetic field. So the kinetic energy term
of the electric dipole superconductor can be written as:
{\large$\frac{|(\hat{\bold p} + \bold p_0\times \bold B)\psi(\bold
r)|^2}{2m^*}$}.

\vspace{5mm}

\begin{center}
\textbf{Supplementary Note 3: The derivation of the Ginzburg-Landau-type equations of
the electric dipole superconductor}
\end{center}

First of all, we minimize the free energy shown in Eq. (5) in
the main text with respect to the complex conjugate of the order
parameter $\psi^*$. For the second and third terms of Eq. (5)
in the main text, we have
\begin{eqnarray}\label{eq:s5}
\delta\int_V d\bold r\left\lbrace\alpha(T)|\psi(\bold r)|^2+\frac{\beta(T)}{2}|\psi(\bold r)|^4\right\rbrace =\int_V d\bold r \left\lbrace\left[\alpha(T)\psi(\bold r)+\beta(T) |\psi(\bold r)|^2\psi(\bold r)\right]\delta \psi^*(\bold r)\right\rbrace.
\end{eqnarray}
For the fourth term, we get
\begin{eqnarray}\label{eq:s6}
\delta\int_V d\bold r\frac{|(\hat {\bold p} + \bold p_0\times \bold B)\psi(\bold r)|^2}{2m^*}&=&\int_V d\bold r\left\lbrace[\frac{i\hbar}{2 m^*}\nabla\delta \psi^*(\bold r)]\cdot[(-i\hbar\nabla+\bold p_0\times\bold B)\psi(\bold r)]\right.\nonumber\\&&\left.+\frac{1}{2 m^*}[(\bold p_0\times\bold B)\delta \psi^*(\bold r)]\cdot[(-i\hbar\nabla+\bold p_0\times\bold B)\psi(\bold r)]\right\rbrace.
\end{eqnarray}
It should be noted that
\begin{eqnarray}\label{eq:s7}
\int_Vd\bold r\left\lbrace\frac{i\hbar}{2 m^*}\nabla\delta \psi^*(\bold r)
 \cdot [(-i\hbar\nabla+\bold p_0\times\bold B)\psi(\bold r)]\right\rbrace &=&\frac{i\hbar}{2 m^*}\oint d\bold S\cdot\lbrace\delta \psi^*(\bold r)(-i\hbar\nabla+\bold p_0\times\bold B)\psi(\bold r)\rbrace \nonumber\\& &-\frac{i\hbar}{2 m^*}\int_V d\bold r\lbrace\delta \psi^*\nabla\cdot[(-i\hbar\nabla+\bold p_0\times\bold B)\psi(\bold r)]\rbrace.
\end{eqnarray}
Combining Eqs. (\ref{eq:s5}), (\ref{eq:s6}) and (\ref{eq:s7}),
we can obtain:
\begin{eqnarray}\label{eq:s8}
\alpha(T)\psi(\bold r)+\beta(T)|\psi(\bold r)|^2\psi(\bold r)
 + \frac{[\hat{\bold p}+\bold p_0\times \bold B]^2\psi(\bold r)}{2m^*}=0 ,
\end{eqnarray}
and
\begin{eqnarray}\label{eq:s9}
[-i\hbar\nabla+\bold p_0\times\bold B]_n\psi(\bold r)=0.
\end{eqnarray}

Eq. (\ref{eq:s8}) is the first Ginzburg-Landau-type equation
and Eq. (\ref{eq:s9}) is the boundary condition for the first
Ginzburg-Landau-type equation, where the subscript n stands for the
component perpendicular to the surface. Here, we emphasize that this
boundary condition is actually the requirement of the variational
principle. In fact, if we substitute Eq. (\ref{eq:s9}) into
Eq. (8) in the main text, we can get $\bold {J_p}_n=0$, which
means that there is no electric dipole current entering or leaving
the electric dipole superconductor. Similar discussions on boundary
condition of the (monopole) superconductor can be found in the
original paper written by Ginzburg and Landau\cite{GLequation}.

Next, we minimize the free energy with respect to the vector
potential $\bold A$. For the fourth term of Eq. (5) in the
main text, we have
\begin{eqnarray}\label{eq:s10}
\delta\int_Vd\bold r\frac{|(\hat{\bold p} + \bold p_0\times \bold B)\psi(\bold r)|^2}{2m^*} &
=&\frac{1}{2m^{\ast}}\int_V d\bold r[\bold p_0\times\delta(\nabla\times\bold A)]\psi\cdot[(i\hbar\nabla
 +\bold p_0\times(\nabla\times\bold A))\psi^*]+c.c \nonumber\\
 &=&\frac{1}{2m^{\ast}}\int_V d\bold r\delta(\nabla\times\bold A)\cdot \lbrace [\psi(i\hbar\nabla
 +\bold p_0\times(\nabla\times \bold A))\psi^*]\times\bold p_0\rbrace+c.c \nonumber\\
 &=&\frac{1}{2m^{\ast}}\int_V\,d\bold r\,\delta \bold A\cdot\lbrace\nabla\times\lbrace[\psi(i\hbar\nabla
 +\bold p_0\times(\nabla\times \bold A))\psi^*]\times\bold p_0\rbrace\rbrace \nonumber\\
&&\left.+\frac{1}{2m^{\ast}}\oint d\bold S\cdot\lbrace \delta\bold A\times[[\psi(i\hbar\nabla\right.
 +\bold p_0\times(\nabla\times \bold A))\psi^*]\times\bold p_0]\rbrace+c.c.
\end{eqnarray}

If we variate the last term of Eq. (5) in the main text with
respect to vector $\bold A$, we get
\begin{eqnarray}\label{eq:s11}
\delta\int_V d\bold r\frac{|\nabla\times\bold A |^2}{2\mu_0}
=\int_V\, d\bold r\delta \bold A\cdot\frac{\nabla\times(\nabla\times\bold A)}{\mu_0}
+\oint d\bold S\cdot\frac{[\delta \bold A\times(\nabla\times\bold A)]}{\mu_0}.
\end{eqnarray}

Combining Eqs. (\ref{eq:s10}) and (\ref{eq:s11}), we can get the
second Ginzburg-Landau-type equation, i.e.,
\begin{eqnarray}\label{eq:s12}
\nabla\times\bold B=-\mu_0 \nabla\times(\bold J_p\times \hat p_0) ,
\end{eqnarray}
where
\begin{eqnarray}\label{eq:s13}
\bold J_p \equiv \frac{p_0}{2 m^*}[i\hbar(\psi\nabla\psi^*-\psi^*\nabla\psi)+2\bold p_0\times\bold B\,|\psi|^2].
\end{eqnarray}
It should be noted that in the derivation the surface integral
vanishes due to the requirement of free energy minimization
\cite{gennes1989superconductivity, GLequation}.

\begin{figure}[!htb]
\centering
\includegraphics[height=3.9cm, width=12cm, angle=0]{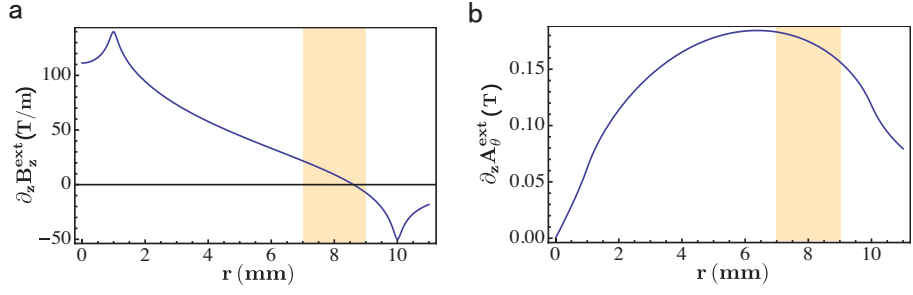} \\
\caption{ \textbf{The spatial variations of the magnetic field and the
vector potential created by the cylindrical hollow conductor.}
(\textbf{a}) The spatial variation of the magnetic field
$\partial_z B_z^{ext}$ created by the cylindrical hollow conductor
versus the coordinate $r$ in the specimen plane $m$. (\textbf{b})
$\partial_z A_{\theta}^{ext}$ versus the coordinate $r$ in the
specimen plane $m$. This magnetic field will be applied on the
electric dipole superconductor specimen. The parameters are same as
in Fig. 2 in the main text.
\label{fig:S1}}
\end{figure}

\vspace{5mm}

\begin{center}
\textbf{Supplementary Note 4: The non-uniform magnetic field created by the electric
current in the cylindrical hollow conductor}
\end{center}

In the article, we consider that a non-uniform magnetic field is
created by a cylindrical hollow conductor. The inner radius of the
cylindrical hollow conductor is $R_{in}$, the outer radius is
$R_{out}$ and the length is $h$ (see Fig. 2a and 2b in the main text).
A uniform electric current density $j$ along the azimuthal direction
is applied in this conductor. In the following, we calculate the
non-uniform magnetic field caused by this electric current in the
cylindrical hollow conductor. This non-uniform magnetic field is the
external magnetic field applied on the electric dipole
superconductor specimen.

To deduce the magnetic field caused by the electric current in the
hollow conductor, we start from the analytical expression of the
magnetic field induced by a circular current loop of radius $R$
carrying current $I$. The circular current loop lies in the $z=z_0$
plane, and is centered at $z$ axis. Then the magnetic field $\bold
B$ and magnetic vector potential $\bold A$ at $(r, \theta, z)$ (in
cylindrical coordinate) can be immediately calculated through
Biot-Savart law\cite{griffiths,simpson}. The results are
\begin{eqnarray}\label{eq:s14}
B_r(r, \theta, z)=&\frac{\mu_0 I (z-z_0)}{2 \pi \xi^2\eta r}[(r^2+R^2
+(z-z_0)^2)E(k)-\xi^2 K(k)],
\end{eqnarray}
\begin{eqnarray}\label{eq:s15}
B_{\theta}(r, \theta, z)=0,
\end{eqnarray}
\begin{eqnarray}\label{eq:s16}
B_z(r, \theta, z)=&-\frac{\mu_0}{2 \pi \xi^2\eta}[(r^2-R^2
+(z-z_0)^2)E(k)-\xi^2 K(k)],
\end{eqnarray}
\begin{eqnarray}\label{eq:s17}
\begin{split}
\partial_z B_z (r, \theta, z)=&\frac{\mu_0 I (z-z_0)}{2 \pi \xi^4\eta^3}[(6 R^2(r^2-(z-z_0)^2)\\
&-7 R^4+(r^2+(z-z_0)^2)^2)E(k)\\
&-\xi^2(r^2-R^2+(z-z_0)^2)K(k)],
\end{split}
\end{eqnarray}
\begin{eqnarray}\label{eq:s18}
\partial_z A_{\theta}(r, \theta, z)=-B_r(r, \theta, z).
\end{eqnarray}
In the above analytical expressions, $\xi=\sqrt{R^2+r^2+(z-z_0)^2-2
r R}$, $\eta=\sqrt{R^2+r^2+(z-z_0)^2+2 r R}$, $k=1-\xi^2/\eta^2$,
and $K(k)$ and $E(k)$ are the first kind and the second kind of the
complete elliptic integral. By using the Biot-Savart law and the
principle of superposition, we calculate the total external magnetic
field gradient $\partial_z B_z^{ext}$ and the gradient of magnetic
vector potential $\partial_z A_{\theta}^{ext}$. It gives
\begin{eqnarray}\label{eq:s19}
 \partial_z B_z^{ext} &= &\frac{\mu_0 j}{2 \pi}\int_{-h/2}^{h/2}dz_0\int_{R_{in}}^{R_{out}}dR
 \left\lbrace\frac{z-z_0}{\xi^4\eta^3}\left\lbrace\left[6 R^2(r^2-(z-z_0)^2)-7 R^4 \right.\right.\right. \nonumber\\
  &&\left.\left.\left. + (r^2+(z-z_0)^2)^2\right]E(k)-\xi^2[r^2-R^2
 +(z-z_0)^2]K(k)\right\rbrace\right\rbrace
\end{eqnarray}
and
\begin{eqnarray}\label{eq:s20}
\partial_z A_{\theta}^{ext}=-B_r=-\frac{\mu_0 j}{2 \pi}\int_{-h/2}^{h/2}dz_0\int_{R_{in}}^{R_{out}}dR \left\lbrace\frac{z-z_0}{\xi^2\eta r}\left[(r^2+R^2+(z-z_0)^2)E(k)-\xi^2 K(k)\right]\right\rbrace.
\end{eqnarray}
Due to the rotational symmetry of the cylindrical hollow conductor,
$\partial_z B_z^{ext}$ and $\partial_z A_{\theta}^{ext}$ are
independent of the angle $\theta$.

Fig. \ref{fig:S1}a and \ref{fig:S1}b in the Supplementary
Note 4 show respectively the spatial variation of the magnetic
field $\partial_z B_z^{ext}$ and $\partial_z A_{\theta}^{ext}$
versus the coordinate $r$ in the specimen plane $m$. Here the
cylindrical hollow conductor sizes are $R_{in}=1mm$, $R_{out}=1\,cm$
and $h=1.5\,cm$. The current density in conductor $j=10^8 A/m^2$. We
can see that $\partial_z B_z^{ext}$ is quite large while $r<R_{in}$.
So we suggest to put the electric dipole superconductor specimen in
the hollow ($r<R_{in}$) if to investigate the Meissner-type effect
of the electric dipole superconductor. In addition, Fig.
\ref{fig:S1}a and \ref{fig:S1}b in the Supplementary Note 4
also show that $\partial_z B_z^{ext}$ is relatively small and
$\partial_z A_{\theta}^{ext}$ is large when below the cylindrical
hollow conductor with $7mm<r<9mm$ (see the yellow color shadow
region), and that is why we choose to put the annular dipole
superconductor specimen in there for the investigation of the
zero electric dipole resistance.

\vspace{5mm}

\begin{center}
\textbf{Supplementary Note 5: The magnetic field induced by the super electric dipole
current}
\end{center}

In this section, we calculate the magnetic field induced by the
super electric dipole current in the bilayer exciton system. In the
bilayer exciton system, the electric dipole current can be viewed as
counter-flow electric currents. Since the bilayer system has the
reflection symmetry about the middle plane $m$ and the rotational
symmetry about the $z$ axis, the induced magnetic field in plane $m$
only has the nonzero r-component $B_r$. Although the $z$-component
of the induced magnetic field vanishes in plane $m$, the spatial
variation $\partial_z B_z^{ind}$ does not vanish. Moreover, we note
that the induced magnetic field $B_r^{ind}$ and its spatial
variation $\partial_z B_z^{ind}$ are twice of that induced by the
electric current in one layer. We will calculate the magnetic field
induced by the electric current in one layer, and then times $2$. At
last, we obtain $\partial_z B_z^{ind}$ and $B_r^{ind}$ in plane $m$:
\begin{eqnarray}\label{eq:s21}
 \partial_z B_z^{ind}(r) &= &\frac{\mu_0}{2 \pi}\int_{0}^{r_{out}}d R
 \left\lbrace\frac{J_{p}}{\xi_0^4\eta_0^3}[(6 R^2(r^2-(d/2)^2)-7 R^4 \right.\nonumber\\
 &&+ \left.(r^2+(d/2)^2)^2)E(k_0)-\xi_0^2(r^2-R^2+(d/2)^2)K(k_0)]\right\rbrace
\end{eqnarray}
\begin{eqnarray}\label{eq:s22}
 B_r^{ind}(r)=-\frac{\mu_0}{2 \pi}\int_0^{r_{out}}d R
\left\lbrace\frac{J_{p}}{\xi_0^2\eta_0
r}[(r^2+R^2+(d/2)^2)E(k_0)-\xi_0^2 K(k_0)]\right\rbrace
\end{eqnarray}
where $\xi_0=\sqrt{R^2+r^2+(d/2)^2-2 r R}$,
$\eta_0=\sqrt{R^2+r^2+(d/2)^2+2 r R}$ and $k_0=1-\xi_0^2/\eta_0^2$.
\\[6mm]

\begin{center}
\textbf{Supplementary References}
\end{center}

\end{widetext}

\end{document}